\documentclass[prletter,showpacs,twocolumn,typeset,superscriptaddress]{revtex4}
\usepackage{amsfonts}
\usepackage{amsmath}
\usepackage{amssymb}
\usepackage{graphicx}
\usepackage{epsfig}
\usepackage{dcolumn}
\usepackage{bm}

\begin{document}
\preprint{APS}

\title{Linearly independent pure-state decomposition and quantum state discrimination}
\author{Luis Roa}
\author{Alejandra Maldonado-Trapp}
\author{Marcelo Alid}
\affiliation{Center for Quantum Optics and Quantum Information, CEFOP,
Departamento de F\'{\i}sica, Universidad de Concepci\'{o}n, Casilla
160-C, Concepci\'{o}n, Chile.}

\date{\today}

\begin{abstract}
We put the pure-state decomposition mathematical property of a mixed state to a physical test.
We begin by characterizing all the possible decompositions of a rank-two mixed state by means of the complex overlap between two involved states.
The physical test proposes a scheme of quantum state recognition of one of the two
linearly independent states which arise from the decomposition.
We find that the two states associated with the balanced pure-state decomposition have the smaller overlap modulus and therefore the
smallest probability of being discriminated conclusively, while in the nonconclusive scheme they have the highest probability of having an error.
In addition, we design an experimental scheme which allows to discriminate conclusively and optimally two nonorthogonal states prepared with different \textit{a priori} probabilities.
Thus, we propose a physical implementation for this {\it linearly independent pure-state decomposition and state discrimination} test
by using twin photons generated in the process of spontaneous parametric down conversion.
The information-state is encoded in one photon polarization state whereas the second single-photon is used for heralded detection.
\end{abstract}

 \maketitle

\section{Introduction}

In quantum information and quantum computing the physical unit of information is a microscopic quantum system and the information is
encoded in its state \cite{Nielsen}.
If the system is isolated, all of its properties are described by a pure state.
The fundamental property of a pure state is that it can always be expressed as a coherent superposition of linearly independent states which,
for instance, gives account of the accurate quantum interference phenomenon \cite{Cohen}.
If the system is not isolated then, in general, it is correlated with an uncontrollable quantum system, usually called environment,
which introduces decoherence to the state of the system \cite{lrgor}.
In this case the effective state of the system can be described by a mixed state which consists of an incoherent superposition of possible states.
In addition, the partial knowledge of the state of a system that belongs to a reservoir also is required to be effectively described by a mixed state.
Whatever be the cause for a mixed state, it has the fundamental property of having an infinite number of decompositions.
There are some properties that do not
depend on the considered decomposition, for instance, eigenvalues, eigenstates, observable-average, purity, and entropy.
However, nonorthogonal quantum state discrimination protocol makes use of a mixed composition of the possible prepared states averaged on their \textit{a priori} probabilities.
Thus, because of the fundamental property of having an infinite number of decompositions one could discriminate not only the
prepared states but also another set of nonorthogonal states belonging to another decomposition of the prepared mix.
On the other hand, the entanglement average for a bipartite system depends on the considered decomposition.
Since the entanglement is understood as the quantification of resources needed to create a given
entangled state, a value of the entanglement of formation of a mixed state is then defined
to be the average entanglement of the pure-state decomposition, minimized over all possible decompositions \cite{Wootters}.
In this form, the mathematical-decomposition property is a nonclassical characteristic \cite{molerr} and becomes of fundamental physical interest.

When the state belongs to a set of linearly independent (LI) and nonorthogonal states,
it can be recognized conclusively with a probability different from zero \cite{Chefles,Peres,ChB,Riis,LR}.
Another strategy allows a minimum-error tolerance for discriminating the state \cite{Helstrom,Sasaki,Clarke}.
In the simplest case the state is in a two-dimensional Hilbert space and only two different states are LI.
In a Hilbert space with dimension higher than two, the states require additional constraints to be LI \cite{petal}.
In order to simplify the study of the addressed problem we have considered a rank-two mixed state.
In this work we relate the LI pure-state decomposition property of a given mixed state to the unambiguous quantum state discrimination (UQSD) protocol for two nonorthogonal states.

\section{Mixed state decomposition}

We consider a quantum system prepared in a mixed state whose spectral decomposition is given by
\begin{equation}
\rho=\lambda_1|\lambda_1\rangle\langle\lambda_1|+\lambda_2|\lambda_2\rangle\langle\lambda_2|,
\label{rho-lambda}
\end{equation}
being $\{|\lambda_1\rangle,|\lambda_2\rangle\}$ the eigenstates and $\{\lambda_1,\lambda_2\}$ the eigenvalues respectively.
We recall that $\lambda_i\in[0,1]$, $\lambda_1+\lambda_2=1$, and $\langle\lambda_1|\lambda_2\rangle=0$.
We shall assume $\lambda_1$ or $\lambda_2$ to be different from zero.
This spectral-decomposition is unique and its purity is completely characterized by the eigenvalues.
However, there are infinite possible \textit{pure-state decompositions} for the same mixed state \cite{Wootters}.
Here we introduce two nonorthogonal states $|\beta_1\rangle$ and $|\beta_2\rangle$ with inner product $\langle\beta_1|\beta_2\rangle=\beta$.
In terms of these states and their biorthogonal ones \cite{Wong} the identity can be represented as follows:
\begin{equation}
I=\frac{|\beta_1\rangle-\beta^*|\beta_2\rangle}{1-|\beta|^2}\langle\beta_1|
+
\frac{|\beta_2\rangle-\beta|\beta_1\rangle}{1-|\beta|^2}\langle\beta_2|.
\label{I-beta}
\end{equation}
This expression becomes the well known canonical one for $\beta=0$.
Making use of the (\ref{I-beta}) identity we can find all the decompositions of $\rho$ whose forms are:
\begin{equation}
\rho=p_1|\beta_1\rangle\langle\beta_1|+p_2|\beta_2\rangle\langle\beta_2|.
\label{rho-beta}
\end{equation}
Here $p_1$ and $p_2$ play the role of the \textit{a priori} probabilities associated with the $|\beta_1\rangle$ and $|\beta_2\rangle$ states, respectively.
After some algebra we obtain
\begin{subequations}
\begin{eqnarray}
p_1&=&\frac{\lambda_1\lambda_2}{\lambda_1+(\lambda_2-\lambda_1)|\gamma|^2}, \label{p1}\\
p_2&=&\frac{\lambda_1^2+(\lambda_2-\lambda_1)|\gamma|^2}{\lambda_1+(\lambda_2-\lambda_1)|\gamma|^2}, \label{p2}
\end{eqnarray}%
\label{p1p2}%
\end{subequations}%
satisfying $0\leq p_i\leq1$ and $p_1+p_2=1$. We note that $p_1$, as a function of $|\gamma|$, is
monotonically increasing (for $\lambda_1<\lambda_2$) or decreasing (for $\lambda_1>\lambda_2$) enclosed by $\lambda_1$ and $\lambda_2$.
The $|\gamma|$-decomposition states become
\begin{subequations}
\begin{eqnarray}
|\beta_1\rangle&=&\gamma|\lambda_1\rangle+\sqrt{1-|\gamma|^2}|\lambda_2\rangle, \label{beta1}\\
|\beta_2\rangle&=&\frac{\lambda_1\sqrt{1-|\gamma|^2}|\lambda_1\rangle-\lambda_2\gamma^{*}|\lambda_2\rangle}
{\sqrt{\lambda_1^2+(\lambda_2-\lambda_1)|\gamma|^2}}, \label{beta2}
\end{eqnarray}%
\label{beta1beta2}%
\end{subequations}%
where the $\gamma=|\gamma|e^{i\theta}$ parameter is the component of the $|\lambda_1\rangle$ eigenstate in the $|\beta_1\rangle$ state,
in short $\gamma=\langle\lambda_1|\beta_1\rangle$, and the phase $\theta$ is the relative phase of the allowed $\{|\beta_i\rangle\}$ states.
From the (\ref{beta1beta2}) expressions we realize that the inner product between the allowed $\{|\beta_i\rangle\}$-decomposition states is given by
\begin{equation}
\langle\beta_1|\beta_2\rangle=\frac{(\lambda_1-\lambda_2)|\gamma|\sqrt{1-|\gamma|^2}}{\sqrt{\lambda_1^2+(\lambda_2-\lambda_1)|\gamma|^2}}e^{-i\theta}.
\label{ppbeta}
\end{equation}
We notice that, as is evident, when $\lambda_1=\lambda_2$ all the possible decompositions are one half of the identity since in this case we get $p_1=p_2$ as well,
and all the possible sets $\{|\beta_i\rangle\}$ are given by the orthogonal states:
\begin{subequations}
\begin{eqnarray}
|\beta_1\rangle&=&|\gamma|e^{-i\theta}|\lambda_1\rangle+\sqrt{1-|\gamma|^2}|\lambda_2\rangle, \label{beta1I}\\
|\beta_2\rangle&=&\sqrt{1-|\gamma|^2}|\lambda_1\rangle-|\gamma|e^{i\theta}|\lambda_2\rangle. \label{beta2I}
\end{eqnarray}%
\label{beta1beta2I}%
\end{subequations}%
From now on we assume $\lambda_1\neq\lambda_2$.
On the other hand, the (\ref{rho-lambda}) spectral-decomposition is recovered for both values $|\gamma|=1$ and $|\gamma|=0$.

The modulus of the overlap (\ref{ppbeta}) is a convex function of $|\gamma|$ being zero for $|\gamma|=0$, $1$ and its maximum value is reached for
\begin{equation}
|\gamma|=\sqrt{\lambda_1}.
\end{equation}
In this case the decomposition corresponds to the \textit{balanced} one since $p_1=p_2=1/2$, and the states become
\begin{subequations}
\begin{eqnarray}
|\beta_1\rangle&=&\sqrt{\lambda_1}e^{-i\theta}|\lambda_1\rangle+\sqrt{\lambda_2}|\lambda_2\rangle, \label{beta1bal}\\
|\beta_2\rangle&=&\sqrt{\lambda_1}|\lambda_1\rangle-\sqrt{\lambda_2}e^{i\theta}|\lambda_2\rangle, \label{beta2bal}
\end{eqnarray}%
\label{beta1beta2bal}%
\end{subequations}%
whose overlap is
\begin{equation}
\langle\beta_1|\beta_2\rangle=(\lambda_1-\lambda_2)e^{i\theta}  \label{beta bal}.
\end{equation}
Thus, the \textit{balanced-decomposition} states have the maximal modulus of the $\langle\beta_1|\beta_2\rangle$ overlap.
In other words, for that decomposition the two states are as close as possible.

\begin{figure}[t]
\includegraphics[angle=360,width=0.40\textwidth]{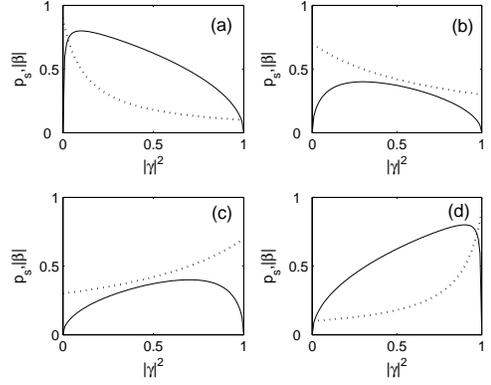}
\caption{Probability $p_1$ (dotted line) and $|\beta|=|\langle\beta_1|\beta_2\rangle|$ (solid line) as functions of $|\gamma|^2$ for different values of $\lambda_1$:
(a) $\lambda_1=0.1$, (b) $\lambda_1=0.3$, (c) $\lambda_1=0.7$, and (d) $\lambda_1=0.9$.} \label{fig1}
\end{figure}

Figure \ref{fig1} shows the $p_1$ probability (solid) and the overlap modulus, $|\beta|$, (dashed) as a function of $|\gamma|^2$ for different values of $\lambda_1$.
We note that $p_1$ is between $\lambda_1$ and $\lambda_2$ and $|\beta|$ reaches its maximal value at $|\gamma|=\sqrt{\lambda_1}$.

By implementing the measurement procedures of two observables, each one with eigenstates $\{|\beta_1\rangle,|\overline{\beta}_1\rangle$,  $\langle\beta_1|\overline{\beta}_1\rangle=0\}$ and $\{|\beta_2\rangle,|\overline{\beta}_2\rangle$,  $\langle\beta_2|\overline{\beta}_2\rangle=0\}$,
we obtain the overlap modulus between the states of the considered decomposition, because of the identity:
\begin{equation}
\langle\beta_1|\rho|\beta_1\rangle+\langle\beta_2|\rho|\beta_2\rangle=1+|\langle\beta_1|\beta_2\rangle|^2.
\end{equation}
Thus, the non-orthogonality of the basis $\{|\beta_1\rangle,|\beta_2\rangle\}$ gives the displacement from $1$ in the square modulus of the overlap $\beta$.

In this section we have characterized by means of the parameter $|\gamma|=|\langle\beta_1|\lambda_1\rangle|$ all possible \textit{pure-state decompositions}
(\ref{rho-beta}) of a given two-rank mixed state (\ref{rho-lambda}).
The modulus of the overlap between the states of the decomposition goes from zero for the spectral-decomposition up to $|\lambda_1-\lambda_2|$ for the balanced-decomposition
whereas its phase is $-\theta$ when $\lambda_1>\lambda_2$ or $-\theta+\pi$ when $\lambda_1<\lambda_2$.
In the next section we relate a possible $|\gamma|$-decomposition of a given density operator to the process for unambiguous nonorthogonal quantum state discrimination.

\section{Quantum state discrimination}

Dieks, Ivanovic, and Peres \cite{Dieks,Ivanovic,Peres} addressed the fundamental problem of discriminating conclusively and unambiguously between two non-orthogonal
states, $|\beta_1\rangle$ and $|\beta_2\rangle$, which are randomly prepared in a quantum system with
\textit{a priori} probabilities $p_1$ and $p_2$ respectively. The optimal success probability for removing the doubt
as to which $|\beta_1\rangle$ or $|\beta_2\rangle$ the system is in, was derived by Jeager
and Shimony \cite{Jeager} obtaining the expressions
\begin{equation}
p_s=1-2\sqrt{p_1p_2}|\beta|
\end{equation}
when $|\beta|\in[0,\min\{\sqrt{p_1/p_2},\sqrt{p_2/p_1}\}]$, and
\begin{equation}
p_s=(1-|\beta|^2)\max\{p_1,p_2\}
\end{equation}
when $|\beta|\in[\min\{\sqrt{p_1/p_2},\sqrt{p_2/p_1}\},1]$.
Replacing in these formulas both $p_1$ and $p_2$ from Eqs. (\ref{p1p2}) and  $|\beta|$ from
Eq. (\ref{ppbeta}) we obtain the optimal probability of success for discriminating unambiguously the two non-orthogonal states of the $|\gamma|-$decomposition of a given $\rho$ mixed state,
\begin{widetext}
\begin{equation}
p_{s}=\left\{
\begin{array}{lll}
1-2\frac{\left\vert \lambda _{1}-\lambda _{2}\right\vert \left\vert \gamma\right\vert\sqrt{\lambda _{1}\lambda _{2}\left( 1-\left\vert \gamma
\right\vert ^{2}\right) }}{\lambda _{1}+\left( \lambda _{2}-\lambda_{1}\right) \left\vert \gamma \right\vert ^{2}}
& \hspace{0.3in}
&\text{if }\hspace{0.1in}0\leq \left\vert\beta\right\vert \leq \min \left\{ \sqrt{\frac{p_{1}}{p_{2}}}\text{ , }\sqrt{\frac{p_{2}}{p_{1}}}\right\},  \\
\left( 1-\frac{\left\vert \lambda _{1}-\lambda _{2}\right\vert^{2}\left\vert \gamma \right\vert ^{2}\left( 1-\left\vert \gamma \right\vert^{2}\right) }
{\lambda _{1}^{2}+\left( \lambda _{2}-\lambda _{1}\right)\left\vert \gamma \right\vert ^{2}}\right) \frac{\max \left\{ \lambda_{1}\lambda _{2}
\text{ , }\lambda _{1}^{2}+\left( \lambda _{2}-\lambda_{1}\right) \left\vert \gamma \right\vert ^{2}\right\} }{\lambda _{1}+\left(
\lambda _{2}-\lambda _{1}\right) \left\vert \gamma \right\vert ^{2}}
& \hspace{0.3in}
& \text{if }\hspace{0.1in}\min \left\{ \sqrt{\frac{p_{1}}{p_{2}}}\text{ , }\sqrt{\frac{p_{2}}{p_{1}}}\right\} \leq \left\vert \beta \right\vert \leq 1.
\end{array}\right.%
\label{ps-p1 p2}
\end{equation}%
\end{widetext}%
As we know, for a given $\lambda_1$ the optimal probability $p_s$ takes its highest value, $1$, for the extreme values $|\gamma|=0$, $1$  which correspond to the spectral-decomposition,
whereas it reaches the smallest value just for $|\gamma|=\sqrt{\lambda_1}$ which corresponds to the balanced-decomposition.
In other words, the states belonging to the balanced-decomposition have the smallest optimal probability of being unambiguously discriminated.
In this case the probability of success becomes $1-|\lambda_1-\lambda_2|$.

\begin{figure}[t]
\includegraphics[angle=360,width=0.40\textwidth]{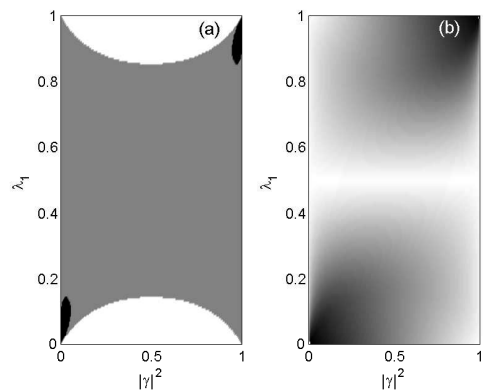}
\caption{(a) regions of the plane ($|\gamma|^2$,$\lambda_1$) where $0\leq|\beta|\leq\min\{\sqrt{p_1/p_2}$,$\sqrt{p_2/p_1}\}$ (gray) and where $\min\{\sqrt{p_1/p_2}$,$\sqrt{p_2/p_1}\}\leq|\beta|\leq1$ (black and white), (b) the $p_s$ optimal probability for unambiguous discriminating of the
$\{|\beta_i\rangle\}$ states as a function of $|\gamma|^2$ and $\lambda_1$. Black color stands for
$p_s=0$, white color means $p_s=1$, and the gray degradation goes linearly from $0$ to $1$.} \label{fig2}
\end{figure}

\begin{figure}[h]
\includegraphics[angle=360,width=0.40\textwidth]{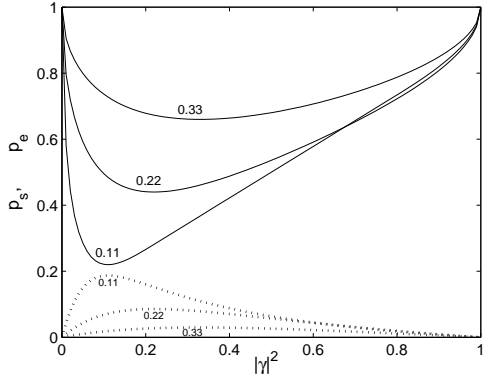}
\caption{Optimal success probability $p_s$ (solid lines) and optimal probability of minimum-error $p_e$ (dotted lines)
as functions of $|\gamma|^2$ for different values of $\lambda_1$, to say: $0.11$,  $0.22$, and $0.33$.
The respective values of $\lambda_1$ are indicated for each of the curves.}
\label{fig3}
\end{figure}

A non-trivial relation between $|\gamma|$ and $\lambda_1$ is obtained from the intervals defined by $|\beta|$ and $\min\{\sqrt{p_1/p_2}$,$\sqrt{p_2/p_1}\}$ in Eq. (\ref{ps-p1 p2}).
Figure \ref{fig2}(a) shows the regions of the ($|\gamma|^2$,$\lambda_1$) plane where $0\leq|\beta|\leq\min\{\sqrt{p_1/p_2}$,$\sqrt{p_2/p_1}\}$ (gray) and where $\min\{\sqrt{p_1/p_2}$,$\sqrt{p_2/p_1}\}\leq|\beta|\leq1$ (black $p_1\leq p_2$ and white $p_1\geq p_2$).
It is worth emphasizing that in the gray area of the Fig. \ref{fig2}(a) both states $|\beta_1\rangle$ and $|\beta_2\rangle$
can be unambiguously discriminated whereas in the white and black zones only the state associated with the higher probability $p_1$ or $p_2$ is discriminated.
Specifically, in the white area only $|\beta_1\rangle$ can be discriminated and in the black one only $|\beta_2\rangle$.
In Fig. \ref{fig2}(b) we plot in degradation black-gray-white the optimal success probability (\ref{ps-p1 p2})
as a function of $|\gamma|^2$ and $\lambda_1$.
In Fig. \ref{fig3} we show the (\ref{ps-p1 p2}) probability (solid lines) as functions of $|\gamma|^2$ for different values of
$\lambda_1$.
Notice that $p_s$ as a function of $|\gamma|^2$ is antisymmetric with respect to $\lambda_1=1/2$ and the minimal values are just at $|\gamma|=\sqrt{\lambda_1}$.

On the other hand, the two states of the $|\gamma|$-decomposition could be recognized tolerating an error.
In this case the strategy of discriminating them with minimum-error
leads to the Helstrom limit \cite{Helstrom}. In other words, the probability of having the smallest error in the identification of the state is
\begin{equation}
p_e=\frac{1}{2}\left(1-\sqrt{1-4p_1p_2|\langle\beta_1|\beta_2\rangle|^2}\right).
\label{Helstrom}
\end{equation}
Replacing in this expression $p_1$ and $p_2$ from Eqs. (\ref{p1p2}) and  $|\langle\beta_1|\beta_2\rangle|$ from
Eq. \ref{ppbeta} we obtain the probability of discriminating with minimum error the $|\gamma|$-decomposition states,
\begin{equation}
p_e=\frac{1}{2}\left(1-\sqrt{1-4\frac{\lambda_1\lambda_2|\lambda_1-\lambda_2|^2|\gamma|^2(1-|\gamma|^2)}
{[\lambda_1+(\lambda_2-\lambda_1)|\gamma|^2]^2}}\right).
\label{pe}
\end{equation}
This probability reaches its smallest value, $0$, in the extreme values $|\gamma|=0$ and $|\gamma|=1$
which correspond to the spectral-decomposition whereas it has the highest value just for $|\gamma|=\sqrt{\lambda_1}$
which corresponds to the balanced-decomposition.
Thus, the states belonging to the balanced-decomposition have the highest probability of discriminating them with minimum-error and this is
$\frac{1}{2}\left(1-\sqrt{1-|\lambda_1-\lambda_2|^2}\right)$.
In Fig. \ref{fig3} we show the (\ref{pe}) probability (dotted lines) as function of $|\gamma|^2$ for different values of
$\lambda_1$. We can see that it has its maximal values for the states of the balanced-decomposition ($|\gamma|=\sqrt{\lambda_1}$) and has the minimum values, $0$,
for the states of the spectral one ($|\gamma|=0$, $1$).

\section{Experimental Scheme for optimal UQSD}

\begin{figure}[h]
\includegraphics[angle=360,width=0.40\textwidth]{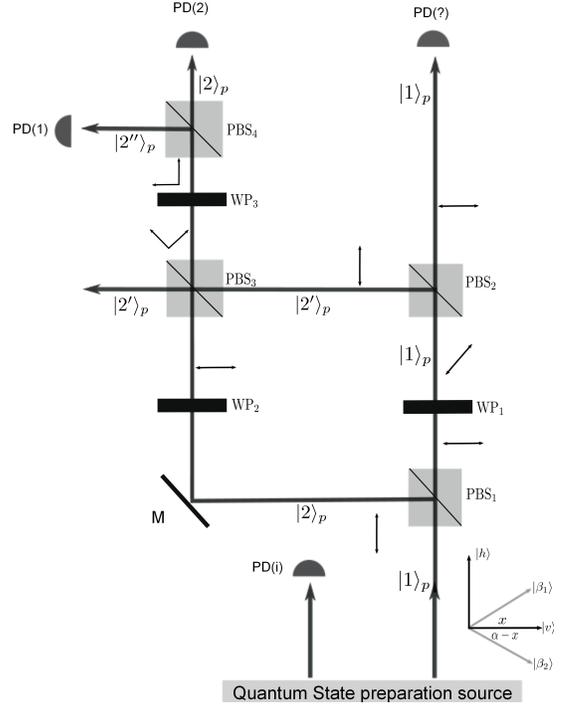}
\caption{Sketch of the experimental setup for conclusively discriminating nonorthogonal quantum states.
We have denoted by $WP$ the wave plate,
by $PBS$ the polarizing beam splitter, by $M$ the mirror, and by $PD$ the single-photon photodiode detectors.} \label{figura5}
\end{figure}

For the unambiguous states discrimination protocol we propose a modified version of the experimental
setup sketched in Ref. \cite{Riis}, see Fig. \ref{figura5}.
We denote by $|h\rangle$ the horizontal
and by $|v\rangle$ the vertical polarization photon states.
For increasing the Hilbert space we consider an ancillary system which consists of a set of four orthogonal effective distinguishable
propagation paths denoted by the states $|1\rangle_p$, $|2\rangle_p$, $|2^\prime\rangle_p$, and $|2^{\prime\prime}\rangle_p$ as shows Fig. \ref{figura5}.

\begin{figure}[t]
\includegraphics[angle=360,width=0.40\textwidth]{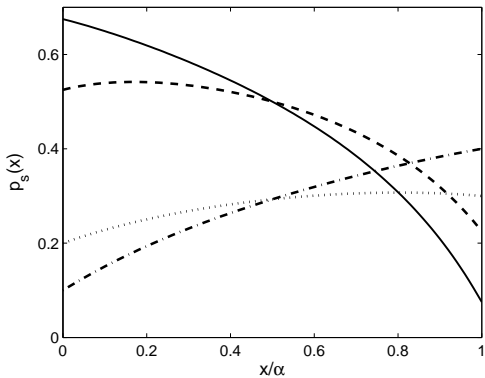}
\caption{Success probability $p_s(x)$ as functions of $x$ for different values of $\alpha$, and $p_1$, say: $\alpha=\pi/3$ with $p_1=0.1$ (solid) and $p_1=0.3$ (dashes),
$\alpha=\pi/4$ with $p_1=0.6$ (dots) and $p_1=0.8$ (dash-dot).}
\label{figpx}
\end{figure}

We assume that the two nonorthogonal possible states $|\beta_i\rangle$, each one having
\textit{a priori} probability $p_i$, enter asymmetrically with respect to the horizontal polarization photon state $|h\rangle$,
specifically,
\begin{subequations}
\begin{eqnarray}
|\beta_1\rangle&=&\cos x|h\rangle+\sin x|v\rangle, \\
|\beta_2\rangle&=&\cos(\alpha-x)|h\rangle-\sin(\alpha-x)|v\rangle.
\end{eqnarray}%
\end{subequations}%
A PBS transmits the horizontal polarization and reflects the vertical one introducing in addition a phase of $\pi/2$.
Thus, after the photon passes through the PBS1 the $|\beta_i\rangle|1\rangle_p$ states are transformed as follows:
\begin{subequations}
\begin{eqnarray}
|\beta_1\rangle|1\rangle_p&\rightarrow &\cos x|h\rangle|1\rangle_p+i\sin x|v\rangle|2\rangle_p, \nonumber\\
|\beta_2\rangle|1\rangle_p&\rightarrow &\cos(\alpha-x)|h\rangle|1\rangle_p-i\sin(\alpha-x)|v\rangle|2\rangle_p.  \nonumber
\end{eqnarray}%
\nonumber%
\end{subequations}%
We consider that the WP1 rotates the photon polarization state $|h\rangle$ an angle $\phi$,
and the WP2 rotates it $\varphi$. Therefore, the states change to
\begin{subequations}
\begin{eqnarray}
|\beta_1\rangle|1\rangle_p&\rightarrow &\cos x(\cos\phi|h\rangle+\sin\phi|v\rangle)|1\rangle_p      \nonumber\\
&& +i\sin x(\sin\varphi|h\rangle-\cos\varphi|v\rangle)|2\rangle_p, \nonumber\\
|\beta_2\rangle|1\rangle_p&\rightarrow &\cos(\alpha-x)(\cos\phi|h\rangle+\sin\phi|v\rangle)|1\rangle_p     \nonumber\\
&& -i\sin(\alpha-x)(\sin\varphi|h\rangle-\cos\varphi|v\rangle)|2\rangle_p.  \nonumber
\end{eqnarray}%
\nonumber%
\end{subequations}%
The unitary effect of the PBS2 on the previous states is
\begin{subequations}
\begin{eqnarray}
|\beta_1\rangle|1\rangle_p&\rightarrow &\cos x(\cos\phi|h\rangle|1\rangle_p+i\sin\phi|v\rangle|2^\prime\rangle_p) \nonumber\\
&& +i\sin x(\sin\varphi|h\rangle-\cos\varphi|v\rangle)|2\rangle_p, \nonumber\\
|\beta_2\rangle|1\rangle_p&\rightarrow &\cos(\alpha-x)(\cos\phi|h\rangle|1\rangle_p+i\sin\phi|v\rangle|2^\prime\rangle_p)   \nonumber\\
&& -i\sin(\alpha-x)(\sin\varphi|h\rangle-\cos\varphi|v\rangle)|2\rangle_p.  \nonumber
\end{eqnarray}%
\nonumber%
\end{subequations}%
Meanwhile the unitary effect of the PBS3 transforms them as follows:
\begin{subequations}
\begin{eqnarray}
|\beta_1\rangle|1\rangle_p &\rightarrow & \cos x\cos\phi|h\rangle|1\rangle_p+i\sqrt{q_{s1}}|\eta_1\rangle|2\rangle_p\nonumber\\
&& +\sin x\cos\varphi|v\rangle|2^\prime\rangle_p, \nonumber\\
|\beta_2\rangle|1\rangle_p &\rightarrow & \cos(\alpha-x)\cos\phi|h\rangle|1\rangle_p-i\sqrt{q_{s2}}|\eta_2\rangle|2\rangle_p \nonumber\\
&& -\sin(\alpha-x)\cos\varphi|v\rangle|2^\prime\rangle_p,  \nonumber
\end{eqnarray}%
\nonumber%
\end{subequations}%
where we have defined the normalized states
\begin{subequations}
\begin{small}
\begin{eqnarray}
|\eta_1\rangle &=& \frac{\sin x\sin\varphi|h\rangle + i\cos x\sin\phi|v\rangle}{\sqrt{q_{s1}}},\\
|\eta_2\rangle &=& \frac{\sin(\alpha-x)\sin\varphi|h\rangle - i\cos(\alpha-x)\sin\phi|v\rangle}{\sqrt{q_{s2}}},
\end{eqnarray}%
\end{small}%
\label{etas}%
\end{subequations}%
and the probabilities
\begin{subequations}
\begin{eqnarray}
q_{s1}&=&\cos^2 x\sin^2\phi+\sin^2 x\sin^2\varphi, \nonumber\\
q_{s2}&=&\cos^2(\alpha-x)\sin^2\phi+\sin^2(\alpha-x)\sin^2\varphi.   \nonumber
\end{eqnarray}%
\nonumber%
\end{subequations}%
From Eqs. (\ref{etas}) we realize that conclusive discrimination can be performed if $|\eta_1\rangle$ and $|\eta_2\rangle$ are orthogonal. This requirement is satisfied when
\begin{equation}
\sin^2\phi=\tan x\tan(\alpha-x)\sin^2\varphi. \label{sinx}
\end{equation}
It is important to point out that the initial angles $\alpha$ and $x$ are restricted in such a way that the right side of Eq. (\ref{sinx})
has to be higher than or equal to $0$ and lower than or equal to $1$. Specifically, it is satisfied for all $\phi$ and $\varphi$ when $0\leq x\leq\alpha$.
Therefore,
by considering satisfied the conditions (\ref{sinx}) and $0\leq x\leq\alpha$, the conclusive discrimination of the nonorthogonal $|\beta_i\rangle$
states becomes just the discrimination between the two orthogonal polarizations $|\eta_1\rangle$ and $|\eta_2\rangle$ of the single photon in the path $|2\rangle_p$.
Replacing
the expression (\ref{sinx}) in the probabilities $q_{si}$ we find, as a function of $x$, the probability $p_s(x)$ of successfully discriminating the $|\beta_i\rangle$ states,
this is
\begin{small}
\begin{eqnarray}
p_s(x)&=&p_1q_{s1}+p_2q_{s2},  \nonumber\\
&=&\left[p_1\frac{\sin x}{\cos(\alpha-x)}+p_2\frac{\sin(\alpha-x)}{\cos x}\right]\sin\alpha\sin^2\varphi.\label{px}
\end{eqnarray}
\end{small}
The first term $p_1q_{s1}$ corresponds to the probability of discriminating the $|\beta_1\rangle$ state and the second term for $|\beta_2\rangle$.
For $x=0$ ($x=\alpha$) there is no probability of discriminating $|\beta_1\rangle$ ($|\beta_2\rangle$).
For other values of $x$ both states can be discriminated with probabilities different from zero.
We can also note that when the initial states $|\beta_i\rangle$ are prepared symmetrically ($x=\alpha/2$) with respect to the horizontal polarization,
the probability (\ref{px}) does not depend on the a priori probabilities $p_1$ and $p_2$.
Therefore, the asymmetry is necessary for the optimization.
Figure \ref{figpx} shows $p_s(x)$ as a function of $x$ for different values of $\alpha$ and $p_1$.
Note that, depending on the values of $\alpha$ and $p_1$, the function $p_s(x)$ has its maximal value inside the interval or at one of the extremes values of $x$, to say: $x=0$ if $p_1<p_2$ or $x=\alpha$ if $p_1>p_2$.
From Eq. (\ref{px}) one analytically finds that the optimal value of the total probability of success, $p_s(x)$, is in $x$ such that
\begin{equation}
\cos x = \frac{\sqrt{p_2}\sin\alpha}{\sqrt{1-2\sqrt{p_1p_2}\cos\alpha}}, \nonumber\\
\end{equation}
and the maximal one becomes
\begin{widetext}
\begin{equation}
p_{s,\max}=\left\{
\begin{array}{lll}
\left(1-2\sqrt{p_1p_2}\cos\alpha\right)\sin^2\varphi & \hspace{0.2in}
& \text{if }\hspace{0.1in} 0\leq\cos\alpha\leq\min\left\{\sqrt{\frac{p_{1}}{p_{2}}}\text{ , }\sqrt{\frac{p_{2}}{p_{1}}}\right\},
\\
\left(1-\cos^2\alpha\right)\max\left\{p_1\text{,}p_2\right\}\sin^2\varphi & \hspace{0.3in}
& \text{if }\hspace{0.1in}\min \left\{ \sqrt{\frac{p_{1}}{p_{2}}}\text{ , }\sqrt{\frac{p_{2}}{p_{1}}}\right\} \leq \cos\alpha \leq 1,
\end{array}\right.%
\label{pdexmax}
\end{equation}%
\end{widetext}%
which is just the well known Jeager and Shimony formula (\ref{ps-p1 p2}) for $\varphi=\pm\pi/2$ (here $\cos\alpha=|\langle\beta_1|\beta_2\rangle|$) \cite{Jeager}.

In this optimal case, the $|\eta_i\rangle$ states of Eqs. (\ref{etas}) become
\begin{subequations}
\begin{eqnarray}
|\eta _{1}\rangle  &=&\cos\xi|h\rangle + \sin\xi|v\rangle, \\
|\eta _{2}\rangle  &=&-\sin\xi|h\rangle+\cos\xi|v\rangle,
\end{eqnarray}%
\label{eta}%
\end{subequations}%
where
\begin{equation}
\cos \xi =\frac{\sqrt{p_1-\sqrt{p_1p_2}\cos\alpha}}{\sqrt{1-2\sqrt{p_1p_2}\cos\alpha}}
\end{equation}
when $0\leq\cos\alpha\leq\min\{\sqrt{p_1/p_2},\sqrt{p_1/p_2}\}$; and
\begin{subequations}
\begin{eqnarray}
|\eta_{1}\rangle  &=& i|v\rangle, \\
|\eta_{2}\rangle  &=& |h\rangle,
\end{eqnarray}%
\label{etau}%
\end{subequations}%
when $\min\{\sqrt{p_1/p_2},\sqrt{p_1/p_2}\}\leq\cos\alpha\leq1$ and $p_1<p_2$; or
\begin{subequations}
\begin{eqnarray}
|\eta_{1}\rangle  &=& |h\rangle, \\
|\eta_{2}\rangle  &=& -i|v\rangle,
\end{eqnarray}%
\label{etauu}%
\end{subequations}%
when $\min\{\sqrt{p_1/p_2},\sqrt{p_1/p_2}\}\leq\cos\alpha\leq1$ and  $p_1>p_2$.

The WP3 (see Fig. \ref{figura5}) rotates the orthogonal photon polarized state (\ref{eta}) in such a way that $|\eta_1\rangle\rightarrow|v\rangle$
and $|\eta_2\rangle\rightarrow|h\rangle$. In this form the PBS4 takes the orthogonal outcome polarization states into the detector PD(1) or PD(2) with optimal probability.
When the process is optimized with respect to $x$ and $\varphi$ ($\varphi=\pm\pi/2$) there is no outcome through the path
$|2^\prime\rangle_p$ and so the inconclusive outcome through the path $|1\rangle_p$ is detected with minimal probability $1-p_{s,\max}$ at the PD(?) photo-detector.
On the other hand, if $\min\{\sqrt{p_1/p_2},\sqrt{p_1/p_2}\}\leq\cos\alpha\leq1$, the $|\eta_i\rangle$ states coincide with the
vertical and the horizonal polarization states as can be seen from Eqs. (\ref{etau}) and (\ref{etauu}); therefore in this case the  WP3 is not required.

Thus, we have designed a physical scheme for discriminating conclusively and optimally two nonorthogonal states associated with different \textit{a priori} probabilities.
Therefore, this designed experimental setup allows one to discriminate desired $|\gamma|$-decomposition states of a two-rank mixed state.

\section{Summary}

In this work we have presented a physical test for the LI pure-state decomposition property of a rank-two mixed state.
We characterized by a complex parameter all the possible LI pure-state decompositions of a mixed state lying in a two-dimensional Hilbert space.
The physical test consists of performing a process of recognition of one of the two linearly independent pure-states which arise from a desired decomposition.
We find that the two states associated with the balanced pure-state decomposition have the
smallest probability of being conclusively discriminated while in the nonconclusive scheme they have the highest probability of having an error.
In addition, we designed an experimental scheme which allows one to discriminate conclusively and optimally two nonorthogonal states prepared with different \textit{a priori} probabilities.
We have proposed an experimental implementation for this \textit{linearly independent pure-state decomposition} and UQSD test
by using a one-photon polarization state generated in the process of spontaneous parametric down conversion (SPDC) where the second single-photon is considered for heralded detection.

For preparing the (\ref{rho-lambda}) state the scheme described in Ref. \cite{Torres} can be implemented.
The signal ($s$) and idler ($i$) twin photons are generated noncollinearly by SPDC in the normalized state
$|\Psi\rangle_{s,i}=\sqrt{\lambda_1}|h\rangle_s|h\rangle_i+\sqrt{\lambda_2}|v\rangle_s|v\rangle_i$.
In this form, by ignoring the polarized state of the idler photon we get the state (\ref{rho-lambda}) for the signal photon with
$|\lambda_1\rangle=|h\rangle$ and $|\lambda_2\rangle=|v\rangle$.
The experiment described in Ref. \cite{Torres} was implemented by using a $351.1$ nm single-mode Ar-ion laser pump with a $200$ mW and $5$-mm-thick BBO crystal,
cut for type-II phase matching which allows a higher stability.
Our proposed scheme for {\it linearly independent pure-state decomposition} and \textit{unambiguous quantum state discrimination} could also be implemented with thit setup.

\begin{acknowledgments}
This work was supported by Grants Basal PFB0824, Milenio ICM P06-067F and FONDECyT N$^{\text{\underline{o}}}$ 1080535.
Two of the authors thank CONICyT-PBCT (A. M.-T.) and CONICyT for scholarship support (M. A.).
\end{acknowledgments}

\end{document}